\documentclass[12pt]{iopart}
\usepackage{bm}
\usepackage{iopams}

\usepackage{graphicx,epsfig}
\newcommand{\text}[1]{\mbox{\scriptsize{#1}}}

\begin{document}
\newtheorem{remark}{{\it Remark}}



\newtheorem{Lemma}{Lemma}
\newtheorem{Theorem}{Theorem}

\hyphenation{super-lat-tice semi-con-ductor}

\title{Nonequilibrium free energy, H theorem and self-sustained oscillations for Boltzmann-BGK 
descriptions of semiconductor superlattices }
\markboth{M Alvaro and L L Bonilla}{H theorem and self-oscillations for Boltzmann-BGK equations}
\author{M \'Alvaro and L L Bonilla}
\address{G. Mill\'an Institute of Fluid Dynamics, Nanoscience and
Industrial Mathematics, Universidad Carlos III de Madrid, Avenida de la Universidad 30, 28911 Legan{\'e}s, Spain }
\eads{\mailto{mariano.alvaro@uc3m.es}, \mailto{bonilla@ing.uc3m.es}}
\date{\today}



\begin{abstract}
Semiconductor superlattices (SL) may be described by a Boltzmann-Poisson kinetic equation with a Bhatnagar-Gross-Krook (BGK) collision term which preserves charge, but not momentum or energy. Under appropriate boundary and voltage bias conditions, these equations exhibit time-periodic oscillations of the current caused by repeated nucleation and motion of charge dipole waves. Despite this clear nonequilibrium behavior, if we `close' the system by attaching insulated contacts to the superlattice and keeping its voltage bias to zero volts, we can prove the H theorem, namely that a free energy $\Phi(t)$ of the kinetic equations is a Lyapunov functional ($\Phi\geq 0$, $d\Phi/dt\leq 0$). Numerical simulations confirm that the free energy decays to its equilibrium value for a closed SL, whereas for an `open' SL under appropriate dc voltage bias and contact conductivity $\Phi(t)$ oscillates in time with the same frequency as the current self-sustained oscillations.
\end{abstract}
\maketitle

\section{Introduction}\label{sec:intro}
In the simplest case, semiconductor superlattices (SL) are artificial one-dimensional crystals comprising many periods, each consisting of two different semiconductors with different gaps but similar lattice constants, e.g. GaAs and AlAs. These nanostructures were invented by Esaki and Tsu in order to develop a device that exhibits Bloch oscillations \cite{esa70}. Electron transport in SL gives rise to a great variety of nonlinear phenomena such as pattern formation, current self-sustained oscillations and chaotic behavior that have been studied theoretically and experimentally \cite{BGr05}. These phenomena are best explored using balance equations for the electron density, current and energy densities in SL minibands and such balance equations can be derived from more complicated Boltzmann type equations in particular hydrodynamic limits. In practice, consistent derivations start from kinetic equations with BGK collision models, named after Bhatnagar, Gross and Krook's simplification of the Boltzmann equation for monoatomic gases and low density plasmas \cite{BGK54}. The idea behind these models is that the distribution function should relax to a local equilibrium distribution whose moments (e.g., density, average velocity and energy) are functions of the corresponding quantities for the instantaneous value of the distribution function. Then the Boltzmann collision term is replaced by a relaxation term proportional to the difference between the distribution function and the local equilibrium function. BGK models are popular because the starting point for deriving balance equations is to find a modified local equilibrium which is the leading order approximation in a perturbation scheme, and this is provided effortlessly by the BGK collision model. 

Kinetic theory descriptions of SL should explain nonequilibrium phenomena such as current self-sustained oscillations that are not common in the kinetic theory of gases. Yet the analysis of these descriptions should be relatively simple because SL are quasi one dimensional systems. Boltzmann type equations for SL go back to the 1970's when Ktitorov, Simin and Sindalovskii (KSS) \cite{KSS72} proposed a relaxation time kinetic equation that included a simple momentum-dissipating but energy-conserving collision term. Later Ignatov and Shashkin \cite{ISh87} replaced KSS's idea of relaxation to global equilibrium by a BGK collision term that preserved charge but dissipated energy, unlike the original charge, momentum and energy conserving BGK collision term \cite{BGK54}. More recently the more complete description including coupling to a Poisson equation was introduced, and balance equations (a nonlinear drift-diffusion equation for the electric field and the electron density coupled to the Poisson equation) derived by using the Chapman-Enskog perturbation method \cite{BEP03}.   Among stable solutions of the balance equations there are self-sustained oscillations of the current through a dc voltage biased SL \cite{BEP03}. The kinetic equation has been solved numerically by a particle method providing confirmation that the balance equations approximate very well the field and electron density including the regime of self-oscillations \cite{CBC09}. Quantum effects \cite{BEs05} and electron transport in SL with two populated minibands have also been considered \cite{BBA08,ABo10}.

In this work, we find the Lyapunov functional for a BGK-Poisson kinetic equation describing a SL for idealized boundary conditions (infinite SL, periodic boundary conditions or finite SL with insulated contacts at zero voltage bias). This functional has the form of a free energy and the corresponding entropy production contains a generalized force which is proportional to the difference between distribution and local equilibrium distribution functions. It is remarkable that formulas of irreversible thermodynamics hold in a regime that is very far from equilibrium. We approximate the free energy functional using the leading order approximation of the Chapman-Enskog method and calculate it numerically in two cases: insulated contacts at zero voltage bias ({\em closed system}) and contacts having finite conductivity at nonzero voltage bias ({\em open system}) in a regime where there are self-sustained oscillations of the current through the SL. In the first case, we have checked that the free energy indeed decreases monotonically towards its equilibrium value, whereas for the case of the {\em open system} (nonzero voltage bias and nonzero contact conductivity) the free energy oscillates in time following the evolution of the current and the electric field inside the SL.

The rest of this paper is as follows. Section \ref{sec:2} presents the model kinetic equation and a generalized drift-diffusion system that can be derived from it by means of the Chapman-Enskog method \cite{BEP03}. In Section \ref{sec:3}, we derive an entropy density functional from a given form of the local equilibrium distribution, and then find a free energy density from which we prove the H theorem provided the contacts at the SL ends are insulating and there is a zero voltage bias between them. In such conditions, any initial condition evolves towards the globally stable equilibrium and the free energy decreases monotonically until it reaches its equilibrium value. Of course under realistic boundary conditions of charge injecting and collecting contacts and nonzero voltage bias, the SL is far from equilibrium and, in fact, self-sustained oscillations of the current due to periodic recycling and motion of charge dipole domains are possible stable solutions \cite{BGr05,CBC09}. In this case, the free energy is no longer a Lyapunov functional and it oscillates periodically in time in the regime of current self-oscillations. In Section \ref{sec:4}, these different regimes and behaviors of the free energy are confirmed by solving numerically the generalized drift-diffusion system, reconstructing the approximate distribution function provided by the Chapman-Enskog method and finding the free energy as a function of time. The last section contains our conclusions.

\section{BPBGK kinetic equation and generalized drift-diffusion balance equation}
\label{sec:2}
In this Section, we present the Boltzmann-Poisson kinetic description of a one-miniband SL and recall the drift-diffusion balance equation that can be derived therefrom by means of the Chapman-Enskog method \cite{BEP03}.

\subsection{Kinetic equation and local equilibrium distribution}
The Boltzmann-Poisson-Bhatnagar-Gross-Krook (BPBGK) system for 1D electron transport in the lowest miniband of a strongly coupled SL is \cite{BEP03}:
\begin{eqnarray} \label{1}
&&{\partial f\over \partial t} + \frac{1}{\hbar}\,\frac{d\mathcal{E}}{d k}\, \frac{\partial f}{\partial x} +\frac{e}{\hbar}\,\frac{\partial W}{\partial x}\,\frac{\partial f}{\partial k} = Q[f]\equiv - \nu_{en}\,  \left(f - f^{FD}\right)-\nu_{imp}\mathcal{A}f , \\
&&\varepsilon\, \frac{\partial^2 W}{\partial x^2} = \frac{e}{ l}\, (n-N_{D}),  \label{2}\\   
&& n(x,t) = \frac{ l}{ 2\pi} \int_{-\pi/l}^{\pi/l} f(x,k,t) dk =
\frac{ l}{ 2\pi} \int_{-\pi/l}^{\pi/l} f^{FD}(k;n(x,t)) dk,\quad \label{3}\\
&& f^{FD}(k;n) = \frac{m^{*}k_{B}T}{\pi\hbar^2}\,\int_{-\infty}^\infty
\ln\left[1+ \exp\left(\frac{\mu - E}{ k_{B}T}\right)\right]\,\frac{\sqrt{2}\,\Gamma^3/
\pi}{ [E-{\cal E}(k)]^4+\Gamma^4}\, dE. \label{4}
\end{eqnarray} 
Here $f$, $\mathcal{A}f=[f(x,k,t-f(x,-k,t)]/2$, $f^{FD}(k;n)$, $n$, $N_{D}$, ${\cal E}(k)$, $d_{B}$, $d_{W}$, $l=d_{B}+d_{W}$, $W$, $\varepsilon$, $m^*$, $k_{B}$, $T$, $\Gamma$, $\nu_{en}$, $\nu_{imp}$ and $-e<0$ are the one-particle distribution function, its odd part (in $k$), the 1D local equilibrium distribution function, the 2D electron density, the 2D doping density, the miniband dispersion relation ($\mathcal{E}(k)$ is even: $\mathcal{E}(-k)=\mathcal{E}(k)$), the barrier width, the well width, the SL period, the electric potential, the SL permittivity, the effective mass of the electron in the lateral directions, the Boltzmann constant, the lattice temperature, the energy broadening of the equilibrium distribution due to collisions \cite{KBa62} (page 28), the constant frequency of the inelastic collisions responsible for energy relaxation, the constant frequency of the elastic impurity collisions and the electron charge, respectively. The chemical potential $\mu$ is a function of the electron density $n$ that can be obtained by inserting (\ref{4}) into (\ref{3}) and solving for $\mu=\mu(n)$. Thus $\mu$ is a functional of $f$ which we may write as $\mu[f]$. 

Integrating (\ref{1}) over $k$ and using $\mu[f]$ given by (\ref{3}), we find the charge continuity equation:
\begin{eqnarray} \label{5}
\frac{e}{l}\,\frac{\partial n}{ \partial t} + \frac{\partial J_n}{\partial x}=0,
\end{eqnarray}
where $J_n(x,t)$ is the electron current density:
\begin{eqnarray}
J_n(x,t) = { e\over 2\pi} \int_{-\pi/l}^{\pi/l} v(k)\, f(x,k,t) dk , \quad v(k)=\frac{1}{\hbar}\,\frac{d\mathcal{E}}{d k} , \label{6}
\end{eqnarray}
and $v(k)$ is the electron group velocity. From (\ref{2}) and (\ref{5}), we obtain
\begin{eqnarray} \label{7}
\varepsilon\,\frac{\partial F}{ \partial t} + J_n =J(t), \quad F=\frac{\partial W}{\partial x},
\end{eqnarray}
where $-F$ is the electric field, $J(t)$ is the total current density and (\ref{7}) is a form of Amp\`ere's law. The main idea behind the BGK collision model is to substitute the linear, quadratic or quartic Boltzmann collision terms, which are {\em nonlocal} in $k$, by a nonlinear collision term, $Q[f]$, that is {\em local} and preserves charge continuity, as in (\ref{5}). It is convenient to use the following tight binding dispersion relation for the miniband with lowest energy:
\begin{eqnarray}
\mathcal{E}(k)= \frac{\Delta}{2}\, (1-\cos kl), \quad v(k)= \frac{\Delta l}{2\hbar}\,\sin kl. \label{8}
\end{eqnarray}
Eq.\ (\ref{8}) contains the first two harmonics in a Fourier series of the periodic function $\mathcal{E}(k)$. (\ref{8}) approximates the band dispersion relation of any crystal provided the band width $\Delta$ is small compared to the energy difference between bands (band gaps) \cite{ash76}. Besides being a reasonable approximation for a SL with well-separated minibands, (\ref{8}) produces simple analytic expressions for the balance equations and is often used in comparisons between theory and experiments \cite{BGr05}.

 The system (\ref{1})-(\ref{4}) is the semiclassical limit of the Wigner system of equations considered in Section 2 of Ref. \cite{BBA08}. Note that the 1D Fermi-Dirac local equilibrium distribution (\ref{4}) can be written as
\begin{eqnarray}
&& f^{FD}(k;n) = \gamma(E[f]), \quad E[f]= \frac{\mathcal{E}(k)-\mu[f]}{k_B T}, \label{9}\\
&& \gamma(E)= \frac{m^{*}k_{B}T}{\pi\hbar^2}\,\int_{-\infty}^\infty\ln(1+ e^{-s})\, \frac{\frac{\sqrt{2}}{\pi}\left(\frac{\Gamma}{k_BT}\right)^3}{(s-E)^4+\left(\frac{\Gamma}{k_BT}\right)^4}\, ds. \label{10}
\end{eqnarray}
$\gamma(E)$ is a decreasing function that takes on positive values for real values of $E$. If broadening due to scattering is negligible, $\Gamma\to 0$, and (\ref{10}) becomes
\begin{eqnarray}
 \gamma(E)= \frac{m^{*}k_{B}T}{\pi\hbar^2}\,\ln(1+ e^{-E}). \label{11}
\end{eqnarray}
In the Boltzmann limit, $E\to \infty$, and we have
\begin{eqnarray}
 \gamma(E)= \frac{m^{*}k_{B}T}{\pi\hbar^2}\, e^{-E}. \label{12}
\end{eqnarray}
This local equilibrium distribution was used by Ignatov and Shashkin \cite{ISh87} to analyze particular solutions of the BPBGK system for constant electric field.

\subsection{Chapman-Enskog perturbation method and drift-diffusion system}
In the hyperbolic limit where collisions and field-dependent terms in the BPBGK system dominate all other terms, it is possible to derive a generalized drift-diffusion equation for the electric field, $-F=-\partial W/\partial x$. Collision and field-dependent terms in (\ref{1}) are of the same order provided $e[F]l/(\hbar)=\nu_{en}$, where $[F]$ gives the order of magnitude of the field. Then $[F]=\hbar\nu_{en}/(el)$. The Poisson equation implies that the distance $[x]$ over which field varies an amount $[F]$ is proportional to $\varepsilon[F]$ divided by $eN_D/l$, so that $[x]=\varepsilon[F]l/(eN_D)=\varepsilon\hbar\nu_{en}/(e^2N_D)$. Let $v_M$ be the order of magnitude of the electron velocity (see below). The reciprocal of the electron residence time is $v_M/[x]$ and the condition that the field-dependent and collision terms dominate all others in (\ref{1}) is:
\begin{eqnarray}
\frac{v_M}{[x]}\ll \nu_{en}\Longrightarrow \delta= \frac{e^2N_Dv_M}{\varepsilon \hbar\nu^2_{en}}\ll 1.\label{13}
\end{eqnarray}
Provided this condition and (\ref{8}) hold, the terms $\partial f/\partial t$ and $v(k)\,\partial f/\partial x$, with $\hbar v(k)= \mathcal{E}'(k)\equiv d\mathcal{E}/dk$, are both of order $\delta\ll 1$ compared to $e\hbar^{-1}\partial W/\partial x\,\partial f/\partial k$ and $Q[f]$ in (\ref{1}). Ignoring these small terms, (\ref{1}) can be approximated by 
\begin{eqnarray}
{\cal L} f^{(0)}(k;F) \equiv \left(\frac{eF}{\hbar}\frac{\partial}{\partial k} + \nu_{en}+ \nu_{imp}\mathcal{A}\right) f^{(0)}(k;F)=\nu_{en}f^{FD}. \label{26}
\end{eqnarray}
The solution of this equation,
\begin{eqnarray}
f^{(0)}(k;F) = \sum_{j=-\infty}^{\infty} {(1-ij {\cal F}/\tau_e)\,f^{FD}_{j}
\over 1 + j^2 {\cal F}^{2}}\, e^{ijkl}, \quad \mathcal{F}=\frac{F}{F_M},\label{24}\\
f^{FD}_{j} =\frac{l}{2\pi}\int_{-\pi/l}^{\pi/l} e^{-ijkl} f^{FD}(k;n) dk, \quad F_M=\frac{\hbar\nu_{en}\tau_{e}}{el}, \,\tau_e= \sqrt{ 1+\frac{\nu_{imp}}{\nu_{en}} },\label{25}
\end{eqnarray}
is the Fermi-Dirac local equilibrium modified by a high electric field. The field-dependent local equilibrium $f^{(0)}$ is the leading order term of a Chapman-Enskog expansion
\begin{eqnarray}
&& f\sim f^{(0)}(k;F) + f^{(1)}(k;F)+ f^{(2)}(k;F), \label{22}\\
&& \varepsilon\,\frac{\partial F}{\partial t} + \mathcal{J}^{(0)}[F] +\mathcal{J}^{(1)}[F] \sim J(t). \label{23}
\end{eqnarray}
(\ref{22}) becomes a power series in the small dimensionless parameter $\delta$ of (\ref{13}) if we nondimensionalize the BPBGK system using the scales $\mathcal{F}$, $\tilde{x}=x/[x]$, $\tilde{k}=kl$, $\tilde{t}=v_Mt/[x]$, etc. The corrections $f^{(m)}$, $m=1,2$, solve the equations
\begin{eqnarray}
{\cal L} f^{(1)} &=&  \left.  - {\partial  f^{(0)}\over \partial t}\right|_{0} - v(k) \frac{\partial f^{(0)}}{\partial x},  \label{27}\\
 {\cal L} f^{(2)} &=&  \left.  - {\partial  f^{(1)}\over \partial t}\right|_{0}
 - v(k) \frac{\partial f^{(1)}}{\partial x} \;
 - \left. {\partial \over \partial t} f^{(0)}\right|_{1}, \quad \label{28}
\end{eqnarray}
in which the subscripts 0 and 1 in their right hand side (RHS) mean that $\varepsilon\, \partial F/\partial t$  is replaced by $J - \mathcal{J}^{(0)}[F]$ and by $-\mathcal{J}^{(1)}[F]$, respectively. The conditions that $f^{(1)}$ and $f^{(2)}$ be bounded and $2\pi/l$ periodic in $k$ determine the functionals $\mathcal{J}^{(m)}[F]$ \cite{BBA08}. When these functionals are inserted in (\ref{23}), we obtain the following generalized drift-diffusion system for the field \cite{BEP03,BBA08}
\begin{eqnarray}
&&\varepsilon {\partial F\over\partial t} + {e v_{M}\over l}\,
{\cal N}\left(F,{\partial F\over\partial x}\right)
=  \varepsilon\,  D\left(F,{\partial F\over\partial x}\right) {\partial^2 F\over \partial x^2}
 + A\left(F,{\partial F\over\partial x}\right) J(t) , \qquad \label{14} \\
&& n=N_D+\frac{\varepsilon l}{e}\frac{\partial F}{\partial x},\quad
{\cal N}  = n V {\cal M}_1 A - {e\Delta l^2\over\hbar^2 \nu^2_{en}\tau_e^2} {B\over 1+{\cal F}^2} \frac{\partial F}{\partial x}, \label{15}\\
&& V({\cal F}) = {2{\cal F}\over 1 + {\cal F}^2} , \quad v_{M} = {\Delta l\, {\cal I}_{1}(M)\over 4\hbar\tau_e {\cal I}_{0}(M)}, \quad \tilde{\mu}=\frac{\mu}{k_BT},\label{16}\\ 
&& A  = 1 + \frac{2 e^2 v_{M}[1-(1+2\tau^2_{e})\mathcal{F}^2]}{ \varepsilon\hbar\nu_{en}^2\tau_{e}^3 (1+ {\cal F}^{2})^3 }\, n {\cal M}_{1}, \,
{\cal M}_m(n/N_D) = { {\cal I}_{m}(\tilde{\mu}) 
{\cal I}_{0}(M)\over {\cal I}_{1}(M) {\cal I}_{0}(\tilde{\mu}) }, \label{17}\\
&&{\cal I}_{m}(\tilde{\mu}) = \frac{l}{2\pi}\int_{-\pi}^{\pi}\gamma\left( \frac{\mathcal{E}(k)}{k_BT}-\tilde{\mu}\right)\cos (m kl)\, dk, \label{18}\\
\label{19} 
&& B = {n{\cal M}_{2}{\cal F} (5-4{\cal F}^2)
\over (1+4{\cal F}^2)^2} - {4\hbar v_{M}(1+\tau_e^2)\over \Delta l\tau_e} n{\cal M}_1(n{\cal M}_1)' {\cal F}{1-{\cal F}^{2}\over (1+ {\cal F}^2)^3} ,\\
&& D = {\Delta^2 l^2 \over 8\hbar^2\nu_{en}\tau^2_{e} (1 + {\cal F}^{2}) }
 \left( 1- {4\hbar v_{M}\tau_e C\over\Delta l} \right), \quad \label{20}\\
&& C= (n {\cal M}_{2})' \frac{1-2\mathcal{F}^2}{1 +  4 {\cal F}^2 } + \frac{8(1+\tau^2_e)\hbar v_{M}}{ \Delta l\tau_e} \left(  {(n {\cal M}_1)'{\cal F}\over 1 + {\cal F}^{2}}\right)^2 . \label{21}
\end{eqnarray}
In these equations, $\tilde{\mu}=M$ provided $n=N_D$. Equation (\ref{14}) is a form of Amp\`ere's law establishing that the sum of the electron current density and Maxwell's displacement current equals the total current density. Numerical solution of both the kinetic system (\ref{1})-(\ref{4}) and of the drift-diffusion system (\ref{14})-(\ref{21}) show that the later is a very good approximation to the former for realistic parameter ranges \cite{CBC09}.

\section{H theorem and entropy for ideal boundary conditions}
\label{sec:3}
In this section, we find a Lyapunov function for the BPBGK kinetic system assuming that the SL is infinite or that it is finite but it has ideal boundary conditions at the contact regions (zero voltage bias and either periodic boundary conditions or insulating contacts with zero conductivity). The Lyapunov functional has the form of a free energy and, for a local equilibrium of Boltzmann type, it produces formulas that are very close to those of equilibrium thermodynamics.

\subsection{Derivation of  a free energy density}
As in the case of the usual BGK equation in the kinetic theory of gases, let us start by finding an entropy functional \cite{zwa01}. Suppose we have a local equilibrium $\gamma(E)$, with $E=E[f]$, which is a non-negative decreasing function for real values of $E$. $\gamma(E)$ can be found by maximizing the entropy functional
\begin{eqnarray}
\mathcal{S}(x,t)= \frac{1}{2\pi}\int_{-\pi/l}^{\pi/l}dk \int_{0}^{f(x,k,t)}ds\, \gamma^{-1}(s)  \label{h1}
\end{eqnarray}
(where $\gamma^{-1}(s)$ is the inverse function of $\gamma(E)$; $\mathcal{S}$ has units of 3D density), subject to the conditions
\begin{eqnarray}
\frac{l}{2\pi}\int_{-\pi/l}^{\pi/l}dk f(x,k,t)=n(x,t),\label{h2}\\
\frac{l}{2\pi}\int_{-\pi/l}^{\pi/l}dk\, \mathcal{E}(k)\, f(x,k,t)=\mathcal{E}(x,t).\label{h3}
\end{eqnarray}
In fact, taking the extremal of
\begin{eqnarray}
\mathcal{S}(x,t)+\frac{\tilde{\mu}(x,t)}{2\pi}\int_{-\pi/l}^{\pi/l}dk f(x,k,t)-\frac{\tilde{\beta}(x,t)}{2\pi}\int_{-\pi/l}^{\pi/l}dk\, \mathcal{E}(k)\, f(x,k,t),\label{h4}
\end{eqnarray}
we obtain $\gamma^{-1}(f^{FD})=\tilde{\beta}\mathcal{E}(k)-\tilde{\mu}$. Then
\begin{eqnarray}
f^{FD}= \gamma(\tilde{\beta}\mathcal{E}(k)-\tilde{\mu}). \label{h5}
\end{eqnarray}
The Lagrange multipliers $\tilde{\beta}$ and $\tilde{\mu}$ have to be calculated by inserting (\ref{h5}) in (\ref{h2})-(\ref{h3}) and solving these equations for $\tilde{\beta}(x,t)$ and $\tilde{\mu}(x,t)$ in terms of $n(x,t)$ and $\mathcal{E}(x,t)$. The entropy density $-\mathcal{S}$ was introduced by Aoki et al \cite{aok07} for a BGK kinetic equation with charge and energy conserving collision terms representing infinitely extended materials with parabolic bands. If we set $\tilde{\beta}=1/(k_BT)$, $\tilde{\mu}=\mu/(k_BT)$, and, by an appropriate choice of $\mu(x,t)$ with constant $T$, we impose that (\ref{h2}) be still satisfied but not (\ref{h3}), $f^{FD}$ given by (\ref{h5}) becomes the same function of  (\ref{9})-(\ref{10}).

To find the Lyapunov functional, we define a free energy density
\begin{eqnarray}
\eta(x,t)= \upsilon(x,t) - \mathcal{S}(x,t)l=\upsilon(x,t)-\frac{l}{2\pi}\int_{-\pi/l}^{\pi/l}dk \int_{0}^{f(x,k,t)}ds\, \gamma^{-1}(s),   \label{h6}
\end{eqnarray}
where $\upsilon(x,t)$ is a functional of the distribution function to be specified later. Using periodicity in $k$, (\ref{h1}) and (\ref{h6}) yield
\begin{eqnarray}
\frac{\partial}{\partial t}(\eta-\upsilon)&=& -\frac{l}{2\pi}\int_{-\pi/l}^{\pi/l}dk\, \gamma^{-1}(f)\, \frac{\partial f}{\partial t}\nonumber\\
&=&\frac{l}{2\pi}\int_{-\pi/l}^{\pi/l}dk\left[ \nu_{en}[f-\gamma(E[f])]+\nu_{imp}\mathcal{A}f+ v(k)\frac{\partial f}{\partial x}+\frac{eF}{\hbar}\frac{\partial f}{\partial k}\right]\gamma^{-1}(f) \nonumber\\
&=&\frac{l\nu_{en}}{2\pi}\int_{-\pi/l}^{\pi/l}dk\frac{\gamma^{-1}(f)-\gamma^{-1}(\gamma(E[f])) }{f-\gamma(E[f])}\,Ê[f-\gamma(E[f])]^2\nonumber\\
&+&\frac{l\nu_{en}}{2\pi}\int_{-\pi/l}^{\pi/l}dk [f-\gamma(E[f])] E[f] \nonumber\\
&+& \frac{l\nu_{imp}}{8\pi}\int_{-\pi/l}^{\pi/l}dk \frac{\gamma^{-1}(f(k))-\gamma^{-1}(f(-k)) }{f(k)-f(-k)}\,Ê[f(k)-f(-k)]^2 \nonumber\\
&+& \frac{l}{2\pi}\frac{\partial}{\partial x}\int_{-\pi/l}^{\pi/l}dk\, v(k)\int_{0}^{f(x,k,t)}ds\, \gamma^{-1}(s).\label{h7}
\end{eqnarray}
The first term in the RHS of this expression is nonpositive because the integrand is proportional to the derivative of the decreasing function $\gamma^{-1}$ at some intermediate point (mean value theorem). The same argument shows that the third term in the RHS is also nonpositive. This term has been written as indicated using that $\int_{-\pi/l}^{\pi/l} g(k)\mathcal{A}f\, dk= \int_{-\pi/l}^{\pi/l} \mathcal{A}g\,\mathcal{A}f\, dk$ for any function $g(k)$. The fourth term is the $x$ derivative of some function. The second term in the RHS of (\ref{h7}) is zero for energy conserving collision terms \cite{aok07} but this is not the case for the BPBGK system. Let us rewrite this term by using
\begin{eqnarray}
&&\frac{l}{2\pi}\frac{\partial}{\partial t}\int_{-\pi/l}^{\pi/l}dk\, \frac{\mathcal{E}(k)}{k_BT}\, f = -\frac{l\nu_{en}}{2\pi}\int_{-\pi/l}^{\pi/l}dk\, [f-\gamma(E[f])] E[f]\nonumber\\
&&-\frac{l}{2\pi}\int_{-\pi/l}^{\pi/l}dk\, v(k)\frac{\mathcal{E}(k)}{k_BT}\frac{\partial f}{\partial x} +\frac{J_nFl}{k_BT}, \label{h8}
 \end{eqnarray}
 after integration by parts. The last term in the RHS of (\ref{h8}) is
 \begin{eqnarray}
&&\frac{l}{k_BT}\, FJ_n = \frac{l F}{k_BT}\left(J(t)-\varepsilon\frac{\partial F}{\partial t}\right)=\frac{l}{k_BT}\,\frac{\partial(JW)}{\partial x}-\frac{\varepsilon l}{2k_BT}\frac{\partial F^2}{\partial t}. \label{h9}
 \end{eqnarray}
 Then (\ref{h8}) and (\ref{h9}) yield
\begin{eqnarray}
&&\frac{l\nu_{en}}{2\pi}\int_{-\pi/l}^{\pi/l}dk\, [f-\gamma(E[f])] E[f]=-\frac{\partial}{\partial t}\left(\frac{l}{2\pi}\,\int_{-\pi/l}^{\pi/l}dk\,\frac{\mathcal{E}(k)}{k_BT}\, f+ \frac{\varepsilon l F^2}{2k_BT} \right)\nonumber\\
&& \quad\quad+\frac{l}{2\pi k_BT}\,\frac{\partial}{\partial x}\left(2\pi J(t)W -\int_{-\pi/l}^{\pi/l}dk\, v(k)\mathcal{E}(k)f \right). \label{h10}
 \end{eqnarray}
Inserting this expression in (\ref{h7}), we obtain
\begin{eqnarray}
\frac{\partial}{\partial t}(\eta-\upsilon)&=& \frac{l\nu_{en}}{2\pi}\int_{-\pi/l}^{\pi/l}dk\frac{\gamma^{-1}(f)-\gamma^{-1}(\gamma(E[f])) }{f-\gamma(E[f])}\,Ê[f-\gamma(E[f])]^2\nonumber\\
&+& \frac{l\nu_{imp}}{8\pi}\int_{-\pi/l}^{\pi/l}dk \frac{\gamma^{-1}(f(k))-\gamma^{-1}(f(-k)) }{f(k)-f(-k)}\,Ê[f(k)-f(-k)]^2 \nonumber\\
&+& l\,\frac{\partial}{\partial x}\left[\frac{1}{2\pi}\int_{-\pi/l}^{\pi/l}dk\, v(k)\left(\int_{0}^{f(x,k,t)}ds\, \gamma^{-1}(s)-\frac{\mathcal{E}(k)}{k_BT}f\right) +\frac{JW}{k_BT} \right]\nonumber\\
&-&\frac{\partial}{\partial t}\left(\frac{l}{2\pi}\,\int_{-\pi/l}^{\pi/l}dk\, \frac{\mathcal{E}(k)\, f}{k_BT} + \frac{\varepsilon l F^2}{2k_BT} \right).\label{h11}
\end{eqnarray}
We have omitted the dependance on $x$, $t$ in the second term of the RHS of (\ref{h11}) to simplify this formula. Selecting now
\begin{eqnarray}
\upsilon&=&\frac{l}{2\pi k_BT}\left(\pi\varepsilon F^2+ \int_{-\pi/l}^{\pi/l}dk\, \mathcal{E}(k) f\right)+C\label{h12}
\end{eqnarray}
($C$ is a constant to be determined), so that
\begin{eqnarray}
\eta&=&\frac{l}{2\pi k_BT}\left(\pi\varepsilon F^2+ \int_{-\pi/l}^{\pi/l}dk\, \mathcal{E}(k) f\right)\nonumber\\
&-&\frac{l}{2\pi}\int_{-\pi/l}^{\pi/l}dk \int_{0}^{f(x,k,t)}ds\, \gamma^{-1}(s)+C,
\label{h13}
\end{eqnarray}
we find
\begin{eqnarray}
&& \frac{\partial\eta}{\partial t}+\frac{\partial J_\eta}{\partial x} = \sigma_\eta,\label{h14}\\
&& J_\eta=- \frac{l}{2\pi}\int_{-\pi/l}^{\pi/l}dk\, v(k)\left(\int_{0}^{f}ds\, \gamma^{-1}(s)-\frac{\mathcal{E}(k)}{k_BT}f\right) -\frac{JlW}{k_BT} \nonumber\\
&&\quad=- \frac{l}{2\pi}\int_{-\pi/l}^{\pi/l}dk\, v(k) \int_{0}^{f}ds\, [\gamma^{-1}(s)-\gamma^{-1}(f^{FD})]+\frac{\mu J_n l}{ek_BT}-\frac{JlW}{k_BT} ,  \label{h15}\\
&&\sigma_\eta= \frac{l\nu_{en}}{2\pi}\int_{-\pi/l}^{\pi/l}dk\frac{\gamma^{-1}(f)-\gamma^{-1}(\gamma(E[f])) }{f-\gamma(E[f])}\,Ê[f-\gamma(E[f])]^2\nonumber\\
&&\quad + \frac{l\nu_{imp}}{8\pi}\int_{-\pi/l}^{\pi/l}dk \frac{\gamma^{-1}(f(k))-\gamma^{-1}(f(-k)) }{f(k)-f(-k)}\,Ê[f(k)-f(-k)]^2 \leq 0.  \label{h16}
\end{eqnarray}
The free energy production $\sigma_\eta$ in (\ref{h16}) is zero only if $f=\gamma(E[f]) = f^{FD}$ (which is even in $k$). Moreover $\gamma^{-1}(s)$ is decreasing from infinity at $s=0$ and therefore it vanishes at a unique value $s=s_0>0$. This value yields the maximum of the integral $\int_0^f\gamma^{-1}(s) ds$. Let us select
\begin{eqnarray}
C&=& \int_{0}^{s_0}ds\, \gamma^{-1}(s), \label{h17}
\end{eqnarray}
then the free energy density becomes
\begin{eqnarray}
\eta&=&\frac{l}{2\pi k_BT}\left(\pi\varepsilon F^2+ \int_{-\pi/l}^{\pi/l}dk\, \mathcal{E}(k) f\right)\nonumber\\
&-&\frac{l}{2\pi}\int_{-\pi/l}^{\pi/l}dk \int_{s_0}^{f(x,k,t)}ds\, \gamma^{-1}(s)\geq 0. \label{h18}
\end{eqnarray}

\subsection{Boltzmann limit and physical meaning of $\eta(x,t)$}
In the Boltzmann limit, (\ref{17}) yields $\gamma^{-1}(s)=-\ln (s/Q)$, with $Q=m^*k_BT/(\pi\hbar^2)$. Then $-\int_0^f\gamma^{-1}(s)\, ds=f\ln (f/Q)-f$, $s_0=C=Q$ because of (\ref{h17}), and (\ref{h18}) gives
\begin{eqnarray}
\frac{k_BT}{l}\,\eta &=&\frac{1}{2\pi}\int_{-\pi/l}^{\pi/l}dk\, \mathcal{E}(k) f+\frac{\varepsilon F^2}{2} -\frac{k_BT}{l}\left(n-\frac{m^*k_BT}{\pi\hbar^2}\right)\nonumber\\
&-&T\left[-\frac{k_B}{2\pi}\int_{-\pi/l}^{\pi/l} f\ln\left(\frac{\pi\hbar^2 f}{m^*k_BT}\right) dk\, \right]. \label{h19}
\end{eqnarray}
The first two terms in the RHS of (\ref{h19}) are the internal energy density (including the contribution from the electric field) and, except for an additive constant, the two last terms can be written as $-T\, s(x,t)$, the temperature times the entropy density if we define
\begin{eqnarray}
s(x,t) &=&-\frac{k_B}{2\pi}\int_{-\pi/l}^{\pi/l} f\,\ln\left(\frac{\pi\hbar^2 f}{m^*k_BT\exp(1)}\right) dk. \label{h19.1}
\end{eqnarray}
Thus $k_BT\eta/l$ is the free energy density. It is remarkable that the classical expression for the free energy holds exactly for out of equilibrium states described by the BPBGK equation in the Boltzmann limit. For other local equilibrium distributions, the entropy functional (\ref{h1}) does not yield the recognizable extensive form of entropy. In the Boltzmann limit, (\ref{3}) gives
\begin{eqnarray}
\mu&=&k_BT\ln\left(\frac{n}{n_0}\right), \label{h20}\\
n_0&=& \frac{m^*k_BT}{\pi\hbar^2}\frac{l}{2\pi} \int_{-\pi/l}^{\pi/l}e^{-\mathcal{E}(k)/(k_BT)} dk= \frac{m^*k_BT}{\pi\hbar^2}e^{-\Delta/(2k_BT)}I_0\left(\frac{\Delta}{2k_BT}\right),   \label{h21}
\end{eqnarray}
which is the usual logarithmic dependence of the chemical potential on the electron density. The explicit formula in (\ref{h21}), where $I_0(x)$ is a modified Bessel function, is found for the tight binding dispersion relation (\ref{8}). If the distribution function is the local equilibrium (\ref{12}), the free energy density (\ref{h19}) becomes
\begin{eqnarray}
\frac{k_BT}{l}\,\eta^{B} &=&\left(\frac{\Delta}{2}-k_BT\right)\frac{n}{l} +\frac{m^*k^2_BT^2}{\pi\hbar^2l}+\frac{\varepsilon F^2}{2}\nonumber\\
&+&\frac{k_BT}{l}n\ln\left[\frac{\pi\hbar^2n}{m^*k_BT\, I_0\left(\frac{\Delta}{2k_BT}\right)}\right].   \label{h22}
\end{eqnarray}

\subsection{Boundary conditions and free energy as a Lyapunov functional}
For an infinitely long SL under zero voltage bias conditions or for a finite SL with periodic boundary conditions, the free energy flux (\ref{h15}) obeys $J_\eta(L,t)=J_\eta(0,t)$, and (\ref{h14})-(\ref{h16}) and (\ref{h18}) imply that the total free energy,
\begin{eqnarray}
\Phi(t)=\frac{k_BTA}{l}\,\int_0^L \eta(x,t)\, dx,   \label{h23}
\end{eqnarray}
is a Lyapunov functional of the BPBGK system ($\Phi(t)\geq 0$, $d\Phi/dt\leq 0$; $A$ is the area of the SL cross section). Then the local equilibrium with $n=N_D$ and $F=0$ is the globally stable stationary solution. These idealized boundary conditions do not hold for a finite SL under voltage bias and with realistic boundary conditions \cite{CBC09}. We have
\begin{eqnarray}
\frac{d\Phi}{dt}=\frac{k_BTA}{l}\left[\int_0^L \sigma_\eta(x,t)\, dx-J_\eta(L,t)+J_\eta(0,t)\right],   \label{h24}
\end{eqnarray}
which no longer has a definite sign.

What are the boundary conditions used to solve the BPBGK system? For a finite SL of length $L=(N+1)l$ ($N$ is the number of SL periods) and a tight binding dispersion relation, appropriate boundary conditions are \cite{CBC09}
\begin{eqnarray}
\mbox{At $x=0$:}\,\,W=0, \quad f^+=\frac{2 \pi \hbar \sigma_0 F}{e \Delta}-
\frac{f^{(0)}}{\int_{0}^{\frac{\pi}{l}} v(k)f^{(0)} \, dk}
\int_{-\frac{\pi}{l}}^{0} v(k)f^{-} \, dk.  \label{h25}\\
\mbox{At $x=L$:}\,\, W=V, \, f^-=\frac{2 \pi \hbar \sigma_L nF}{e \Delta N_D} -
\frac{f^{(0)}}{\int_{-\frac{\pi}{l}}^0 v(k)f^{(0)} \, dk}
\int_0^{\frac{\pi}{l}} v(k)f^{+} \, dk,  \label{h26}
\end{eqnarray}
where $f^\pm$ means $f(x,k,t)$ for sign$(k)=\pm 1$ and $f^{(0)}(k;F)$ is given by (\ref{24}). Integrating the second relations in (\ref{h25}) and (\ref{h26}) over $k$, we get 
\begin{eqnarray}
J_n(0,t)=\sigma_0 F(0,t),\quad J_n(L,t)=\sigma_L F(L,t)\, \frac{n(L,t)}{N_D}, \label{h27}
\end{eqnarray}
These relations are of the forms: $J_n(0,t)=j_0(F(0,t))$ and $J_n(L,t)=j_L(F(L,t))\, n(L,t)/N_D$ in which the contact currents $j_0(F)$ and $j_L(F)$ have been linearized according to Ohm's law (with contact conductivities $\sigma_0$ and $\sigma_L$, respectively). A derivation of these relations from quantum theory can be found in \cite{BPS00}. They are often used in the literature; cf the review \cite{BGr05}. We prove below that $J_\eta(L,t)-J_\eta(0,t)\geq 0$ provided $\sigma_0=\sigma_L=0$ (insulated contacts) and $V=0$ (zero voltage bias). Then the total free energy, $\Phi(t)$, is a Lyapunov functional of the BPBGK system and the distribution function tends to equilibrium which is the unique globally stable stationary solution of the system.
\bigskip

{\em Proof that $J_\eta(L,t)-J_\eta(0,t)\geq 0$ for zero contact conductivity and zero voltage bias.} (Adapted from Chapter 8 of \cite{CIP94}). In (\ref{h16}), the function
\begin{eqnarray}
C(f)=-\int_{0}^{f}ds\, [\gamma^{-1}(s)-\gamma^{-1}(f^{FD})] ,  \label{h28}
\end{eqnarray}
is convex for non-negative $f$ and it reaches a minimum at $f=f^{FD}$. Then Jensen's inequality implies that
\begin{eqnarray}
&& C\left(\int_0^{\pi/l}\Omega(k',k)f(k')dk'\right)\leq\int_0^{\pi/l}\Omega(k',k)C(f(k'))\,dk',\label{h29}
\end{eqnarray}
at $x=L$, where
\begin{eqnarray}
\Omega(k',k)=\frac{|v(k')|\, f^{(0)}(k)}{\int_{-\pi/l}^0|v(\kappa)|\, f^{(0)}(\kappa)d\kappa}=-\frac{v(k')\, f^{(0)}(k)}{\int_{-\pi/l}^0 v(\kappa)\, f^{(0)}(\kappa)d\kappa},  \label{h30}
\end{eqnarray}
with $k'>0$ and $k<0$. We have omitted the dependence on $x=L$ and $t$ in the distribution functions. In the left hand side of (\ref{h29}),
\begin{eqnarray}
&& \int_0^{\pi/l}\Omega(k',k)f(k')dk'= f(k),\label{h31}
\end{eqnarray}
according to (\ref{h26}) with $\sigma_L=0$. We now substitute (\ref{h31}) in (\ref{h29}), multiply the result by $|v(k)|=-v(k)$, integrate over $k$ in $-\pi/l\leq k\leq 0$, and use that $-\int_{-\pi/l}^0 v(k)\Omega(k',k) dk= v(k')$ due to (\ref{h30}). The result is
\begin{eqnarray}
&& -\int_{-\pi/l}^0v(k)\, C(f(k))\, dk\leq\int_0^{\pi/l}v(k)\, C(f(k))\,dk,\quad\mbox{therefore}\nonumber\\
&&\int_{-\pi/l}^{\pi/l}v(k)\, C(f(k))\,dk \geq 0.  \label{h32}
\end{eqnarray}

Similarly, we can prove that $\int_{-\pi/l}^{\pi/l}v(k)\, C(f(k))\,dk \leq 0$ at $x=0$ if $\sigma_0=0$. Since $J_n=0$ for insulating contacts, (\ref{h15}) implies $J_\eta(L,t)-J_\eta(0,t)\geq -JlV/(k_BT)=0$ for $V=0$ (zero voltage bias). Then (\ref{h24}) and (\ref{h16}) yield $d\Phi/dt\leq 0$.

\section{Numerical results}
\label{sec:4}
In order to calculate numerically the Lyapunov functional $\Phi(t)$, we have first obtained $F(x,t)$, $n(x,t)$ and $J(t)$ by solving (\ref{14})-(\ref{21}) under appropriate dc voltage bias and the boundary conditions (\ref{h27}). From equation (\ref{24}) we have calculated the leading order approximation of $f \sim f^{(0)}$ using the Boltzmann local equilibrium distribution (\ref{12}) (in this case $f_j^B = n\, I_j({\Delta\over 2K_BT})/I_0({\Delta\over 2K_BT})$, where $I_j(x)$ is a modified Bessel function). Then we have obtained the free energy density $\eta$ from (\ref{h19}). Finally, we have calculated the Lyapunov functional $\Phi(t)$ from equation (\ref{h23}). 

In our numerical simulations, we have used the following typical parameters \cite{BGr05}: $d_{B}=1.5$ nm, $d_{W}=9$ nm, $l= d_{B}+d_{W} = 10.5$ nm, $N_{D}= 2.5\times 10^{10}$ cm$^{-2}$, $\nu_{en}= 9\times 10^{12}$ Hz, $\nu_{imp}=0$, $N=80$,  $T=5$ K, $\Delta= 2.6$ meV, $\varepsilon = 12.85\,\varepsilon_0$, $A=100\, \mu$m$^2$. We have selected the following units to present our results graphically: $F_M=\hbar\nu_{en}/(el)= 5.64$ kV/cm, $x_{0} = \varepsilon F_M l /(eN_D) = 16.83 $ nm, $v_M = \Delta l /( 2 \hbar) = 20.46 $ km/s,  $J_0=eN_D v_M/l=7.8$ kA/cm$^2$, $t_{0} = x_0/v_M = 0.82 $ ps, $\eta_{0} = {\varepsilon \nu_{en}^2 \hbar^2 \over 2 k_BTe^2l} = 2.76\times 10^{11}$ cm$^{-2}$, $\Phi_0=k_BTA\,\eta_0\, x_0/l= 190.27$ eV. The chosen unit $\eta_0$ reflects the fact that the most important contribution to the free energy density is the electrostatic one.
\\
\begin{figure}
\begin{center}
\includegraphics[scale=1.0,width=5.5cm,angle=0]{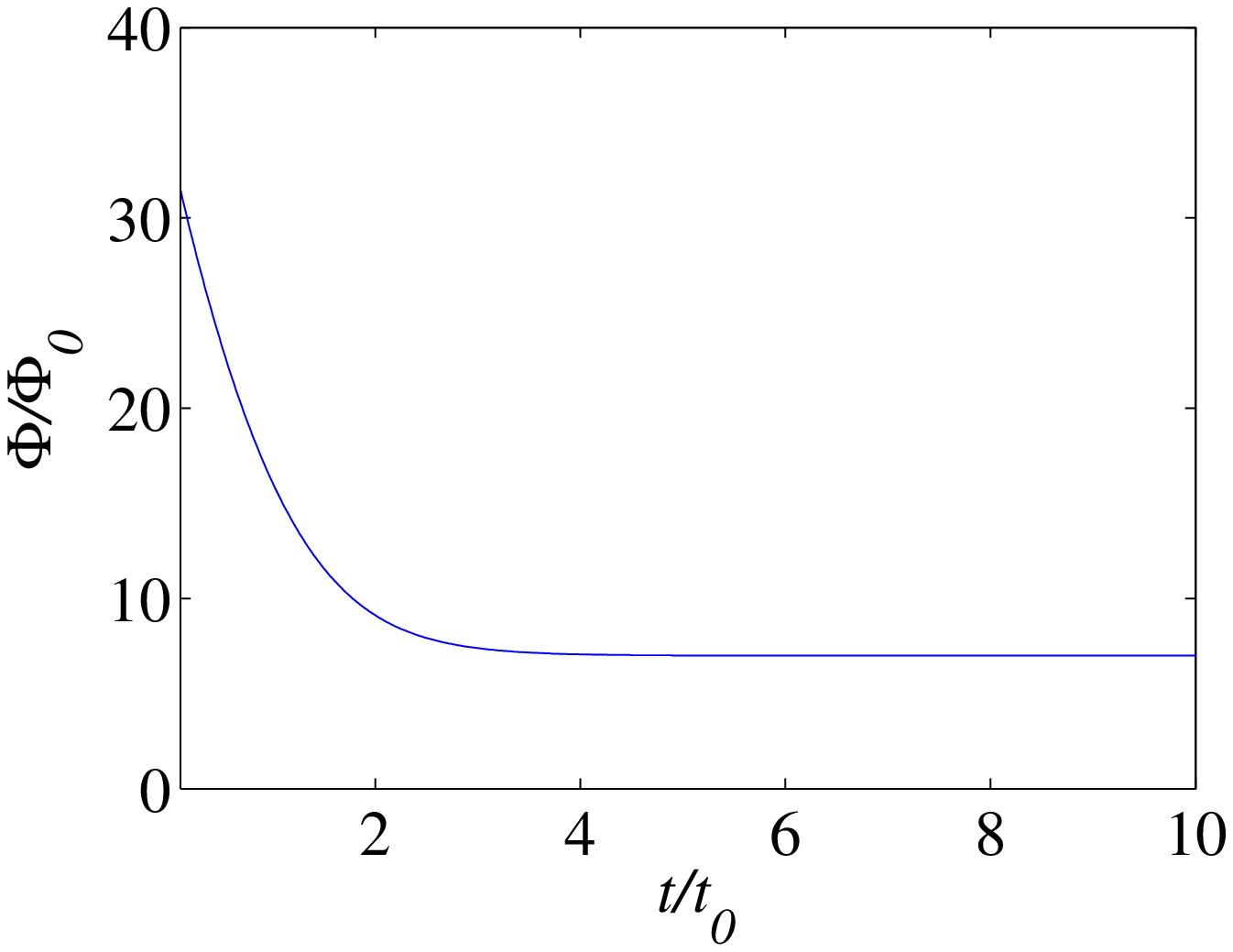} %
\includegraphics[scale=1.0,width=5.5cm,angle=0]{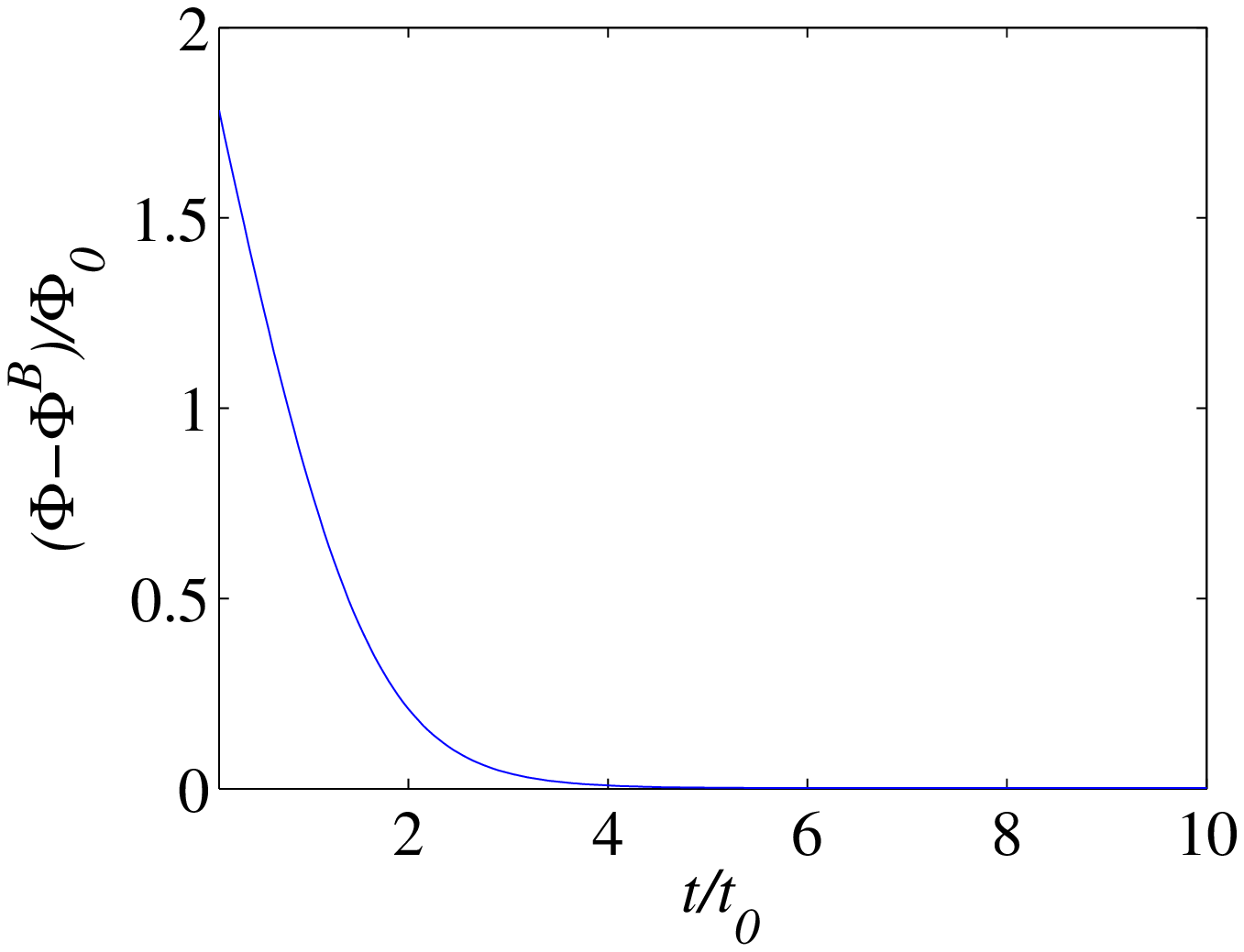} %
\vspace{0.2cm} \caption{Time evolution of the Lyapunov functional $\Phi(t)$ for zero contact conductivity and zero voltage bias. Here $\Phi_0=190.27$ eV, $t_0=0.82$ ps. } \label{fig1}
\end{center}
\end{figure}
\begin{figure}
\begin{center}
\includegraphics[scale=1.0,width=5.5cm,angle=0]{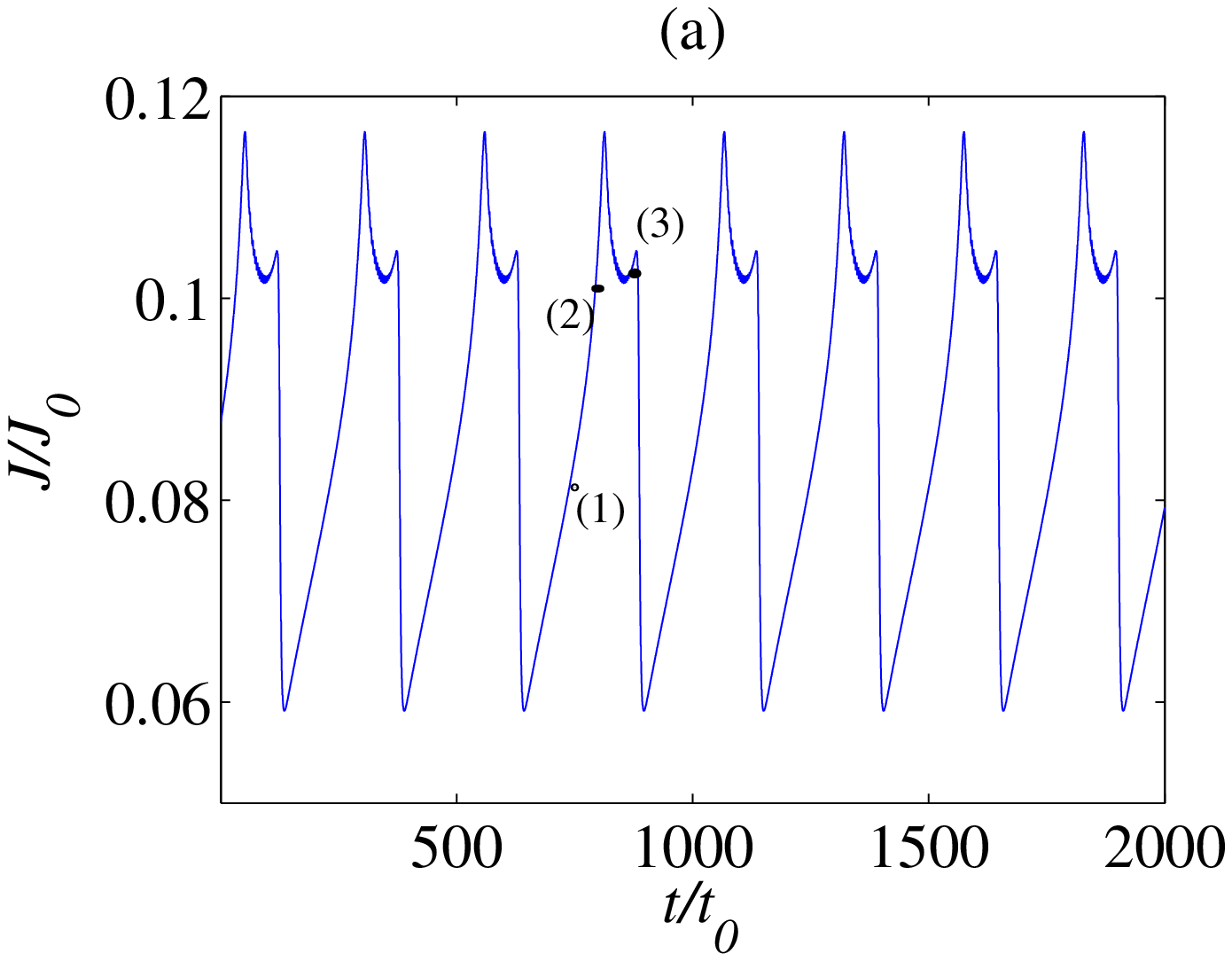} %
\includegraphics[scale=1.0,width=5.5cm,angle=0]{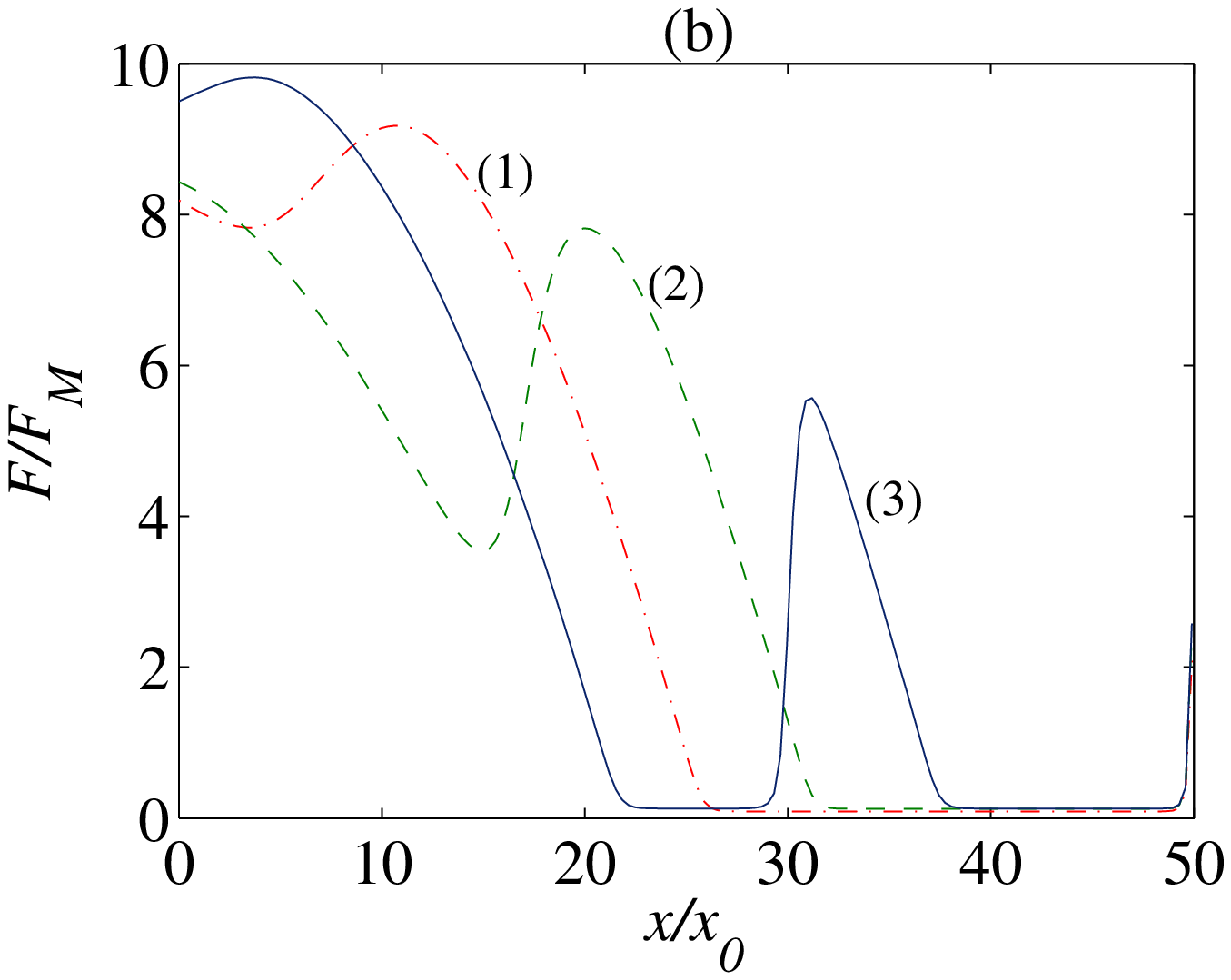} %
\vspace{0.2cm} 
\caption{(a) Total current density vs time. (b) Electric field profiles at several times during one oscillation period. Here $J_0=7.8$ kA/cm$^2$, $t_0=0.82$ ps, $F_M= 5.64$ kV/cm, $x_0=16.83$ nm. } \label{fig2}
\end{center}
\end{figure}
\begin{figure}
\begin{center}
\includegraphics[scale=1.0,width=5.5cm,angle=0]{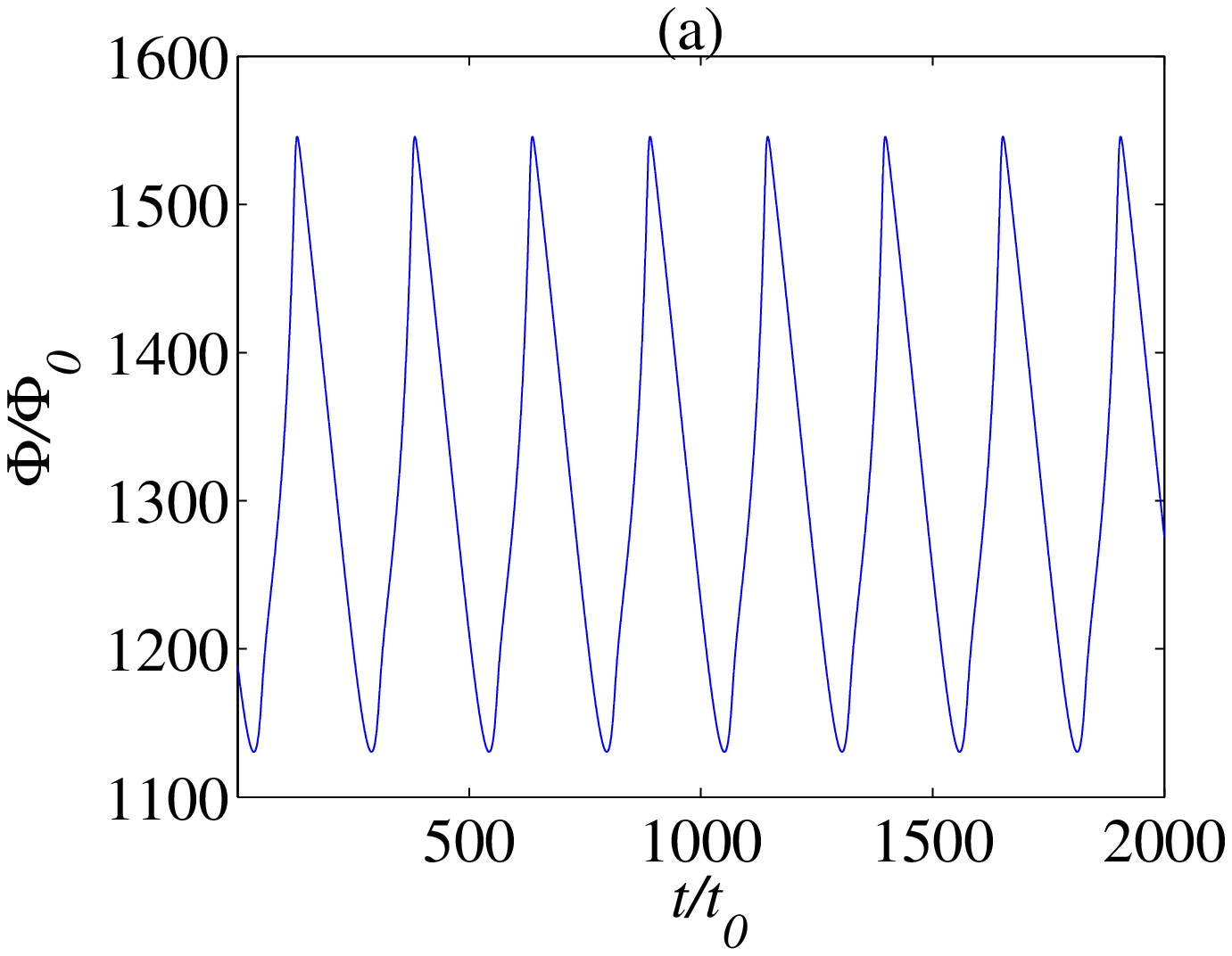} %
\includegraphics[scale=1.0,width=5.5cm,angle=0]{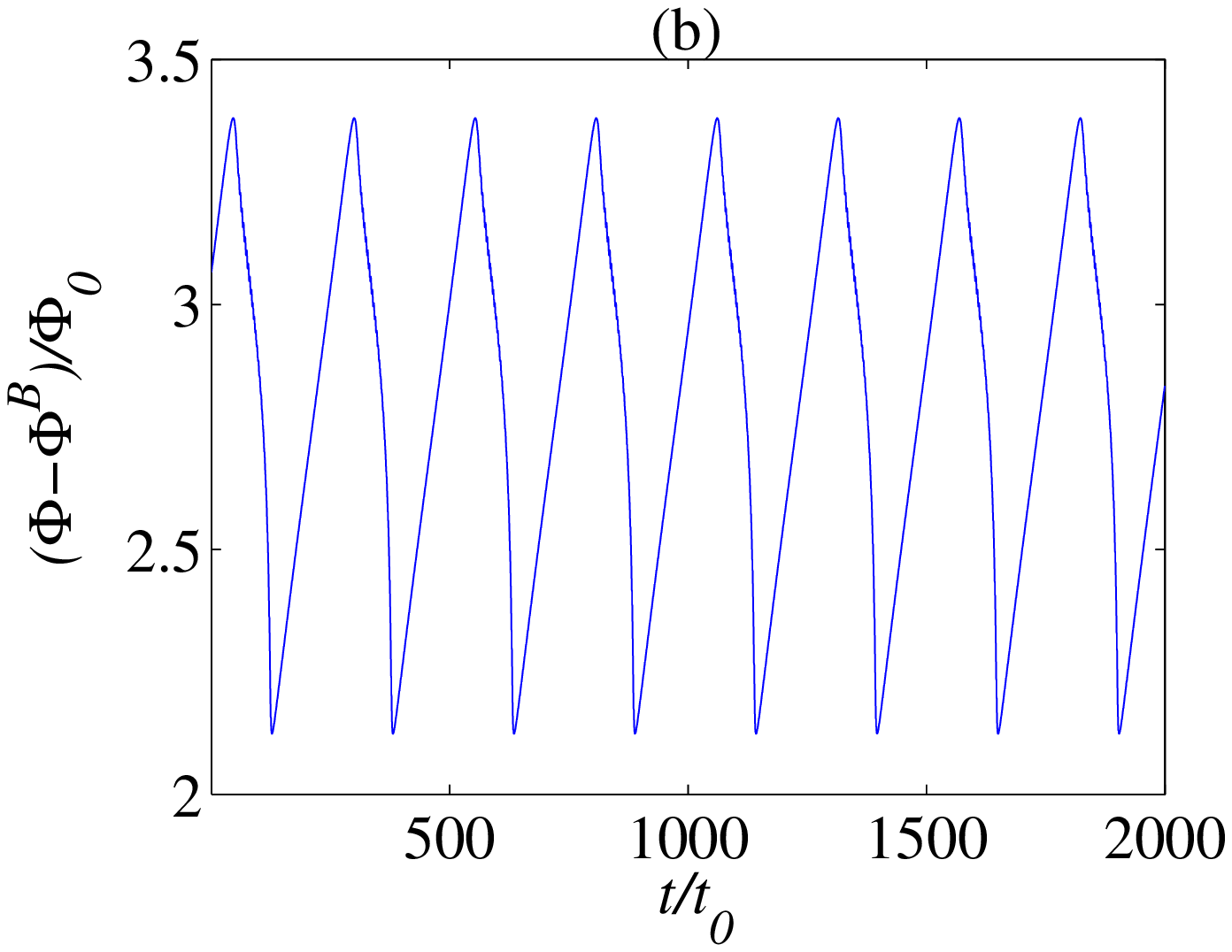} %
\vspace{0.2cm} \caption{(a) Time evolution of the Lyapunov functional $\Phi(t)$ during self sustained current oscillations in a SL with nonzero contact conductivity biased at 1.7 V. (b) Same for $\Phi(t)-\Phi^B(t)$. } \label{fig3}
\end{center}
\end{figure}

We have studied the following cases:
\begin{itemize}
\item Zero contact conductivity and zero voltage bias: We see in figure \ref{fig1}  that the Lyapunov functional $\Phi(t)$ tends asymptotically to the local equilibrium $\Phi^B(t)=\frac{k_BTA}{l}\,\int_0^L \eta^B(x,t)\, dx$, as expected.

\item Realistic contact conditions: For an applied voltage bias $V=1.7$ V and contact conductivities $\sigma_{0}= 1.38\, \Omega^{-1}$m$^{-1}$ and $\sigma_{L}= 0.69\, \Omega^{-1}$m$^{-1}$, the total current exhibits self sustained oscillations due to the repeated nucleation of dipole waves at the injecting contact and their motion towards the collector \cite{BGr05}, as shown in Figure \ref{fig2}. Figure \ref{fig3} shows that free energy $\Phi(t)$ oscillates at the same frequency and, obviously, it is no longer a Lyapunov functional. Global equilibrium is precluded by injection and depletion of carriers at the contacts and the nonzero voltage bias. Figure \ref{fig3}(b) shows that the free energy differs from that of local equilibrium whereas a comparison to Figure \ref{fig3}(a) indicates that $\Phi^B$ has the same order of magnitude as $\Phi$.\\
\end{itemize}

\section{Conclusions}
\label{sec:5}
Electron transport in strongly coupled miniband superlattices is described by a BPBGK kinetic system of equations. For this system, we have found a free energy density $\eta(x,t)$ and its corresponding free energy $\Phi(t)$. We have proved that the free energy is a Lyapunov functional ($\Phi\geq 0$, $d\Phi/dt\leq 0$) for idealized boundary conditions in the SL (zero voltage bias and either zero contact conductivity or periodic boundary conditions). Thus for a {\em closed} SL with insulating contacts and zero voltage bias, the distribution function tends to equilibrium which is the unique globally stable stationary solution of the BPBGK system. In the Boltzmann limit, it is remarkable that the expression of the free energy density $\eta(x,t)$ is similar to that of equilibrium thermodynamics, including the internal energy density minus a $T\, s(x,t)$ term (temperature times the entropy density). However, an {\em open} SL (nonzero voltage and nonzero conductivity at contacts) may be very far from equilibrium and display self-sustained current oscillations due to repeated nucleation and motion of charge dipole waves as in Figure \ref{fig2}. In this case, $d\Phi/dt$ has no definite sign. Numerical solutions of the BPBGK system of equations illustrate the monotonic decay of the free energy of a closed SL towards its equilibrium value and show that, during self-oscillations in an open SL, $\Phi(t)$ oscillates stably and periodically in time 
at the same frequency as the total current.

Our results are based on selecting an entropy density based on the local equilibrium distribution $\gamma(E[f])$ by using the maximum entropy principle and therefore they can be applied to other forms of $\gamma(E)$ (decreasing with $E$, compact support \cite{aok07}) or to higher dimensional models of similar type (collision operator preserving charge but not energy or momentum). 

\ack

This research has been supported by the Spanish Ministerio de Ciencia e Innovaci\'on (MICINN) through Grant FIS2008-04921-C02-01.

\end{document}